\documentclass[twocolumn,prc,aps,superscriptaddress,showpacs]{revtex4}
\usepackage{graphicx}
\usepackage{amssymb}
\usepackage[latin2]{inputenc}
\usepackage[T1]{fontenc}
\begin{document}

\title{Beta-Delayed Proton Emission Branches in $^{43}$Cr}

\date{\today}

\author{M.~Pomorski}
\affiliation{Faculty of Physics, University of Warsaw, 00-681 Warsaw, Poland}
\author{K.~Miernik}
\affiliation{Faculty of Physics, University of Warsaw, 00-681 Warsaw, Poland}
\author{W.~Dominik}
\affiliation{Faculty of Physics, University of Warsaw, 00-681 Warsaw, Poland}
\author{Z.~Janas}
\affiliation{Faculty of Physics, University of Warsaw, 00-681 Warsaw, Poland}
\author{M.~Pf\"utzner}
\email{pfutzner@fuw.edu.pl}
\affiliation{Faculty of Physics, University of Warsaw, 00-681 Warsaw, Poland}
%
%
\author{C.R.~Bingham}
\affiliation{Department of Physics and Astronomy, University of Tennessee, Knoxville, TN 37996, USA}
\author{H.~Czyrkowski}
\affiliation{Faculty of Physics, University of Warsaw, 00-681 Warsaw, Poland}
\author{M.~\'Cwiok}
\affiliation{Faculty of Physics, University of Warsaw, 00-681 Warsaw, Poland}
\author{I.G.~Darby}
\affiliation{Department of Physics and Astronomy, University of Tennessee, Knoxville, TN 37996, USA}
\author{R.~D\k{a}browski}
\affiliation{Faculty of Physics, University of Warsaw, 00-681 Warsaw, Poland}
\author{T.~Ginter}
\affiliation{National Superconducting Cyclotron Laboratory, Michigan State University, East Lansing, MI 48824, USA}
\author{R.~Grzywacz}
\affiliation{Department of Physics and Astronomy, University of Tennessee, Knoxville, TN 37996, USA}
\affiliation{Physics Division, Oak Ridge National Laboratory, Oak Ridge, TN 37831, USA}
\author{M.~Karny}
\affiliation{Faculty of Physics, University of Warsaw, 00-681 Warsaw, Poland}
\author{A.~Korgul}
\affiliation{Faculty of Physics, University of Warsaw, 00-681 Warsaw, Poland}
\author{W.~Ku\'smierz}
\affiliation{Faculty of Physics, University of Warsaw, 00-681 Warsaw, Poland}
\author{S.N.~Liddick}
\affiliation{Department of Physics and Astronomy, University of Tennessee, Knoxville, TN 37996, USA}
\author{M.~Rajabali}
\affiliation{Department of Physics and Astronomy, University of Tennessee, Knoxville, TN 37996, USA}
\author{K.~Rykaczewski}
\affiliation{Physics Division, Oak Ridge National Laboratory, Oak Ridge, TN 37831, USA}
\author{A.~Stolz}
\affiliation{National Superconducting Cyclotron Laboratory, Michigan State University, East Lansing, MI 48824, USA}

\begin{abstract}
The $\beta^+$ decay of very neutron deficient $^{43}$Cr has been studied by means
of an imaging time projection chamber which allowed recording tracks of
charged particles. Events of $\beta$-delayed emission of one-, two-, and three
protons were clearly identified. The absolute branching ratios
for these channels were determined to be ($81 \pm 4$)\%, ($7.1 \pm 0.4$)\%, and
($0.08 \pm 0.03$)\%, respectively. The $^{43}$Cr is thus established as the second
case in which the $\beta$-3\emph{p} decay occurs. Although the feeding to the
proton-bound states in $^{43}$V is expected to be negligible, the large branching
ratio of ($12 \pm 4$)\% for decays without proton emission is found.
\end{abstract}

\maketitle

The $\beta$-delayed emission of protons is a characteristic feature of
very neutron deficient nuclei which have large decay energies and thus substantial
probability of $\beta$ transitions feeding highly excited and particle-unbound
states in daughter nuclei. Since the first observation of delayed proton emission,
almost 50 years ago \cite{Barton}, such decays provided a wealth of information
on structure of neutron-deficient nuclei far from stability, providing tests to nuclear
models and yielding data needed for the understanding of the astrophysical
\emph{rp}-process \cite{BlankBorge}. With the progress in experimental techniques
more exotic systems could be reached and the phenomena of $\beta$-delayed multiparticle
emission become open to investigation. The $\beta$-delayed two-proton
decay was observed for the first time in 1983 in the decays of $^{22}$Al and
$^{26}$P \cite{Cable,Honkanen}. Later, several other cases of such decay
mode were identified \cite{Fynbo1}.

A few years ago, we have communicated
the first observation of $\beta$-delayed three-proton emission, which was found
in the decay of $^{45}$Fe \cite{Miernik1}. The identification of this new
channel was possible due to application of a new type of ionization chamber,
equipped with optical readout --- called the Optical Time Projection Chamber
(OTPC) ---  which was developed to study in detail the
two-proton (\emph{2p}) radioactivity of $^{45}$Fe \cite{Miernik2,Miernik3}.
This detector proved to be extremely sensitive, yielding clear and
unambiguous images of decays of single atoms stopped within its volume.
The evidence of the \emph{3p} channel following $\beta^+$ decay of $^{45}$Fe
was based on four events \cite{Miernik1}. In the same experiment, as a byproduct,
information on the $\beta^+$ decay of $^{43}$Cr was collected as well.
In this Rapid Communication, we present results obtained for the decay
of $^{43}$Cr, which include evidence for the $\beta$-\emph{3p} decay channel.

The first observation of $^{43}$Cr and the first determination on its
decay properties were achieved in 1992 by Borrel et al. using the LISE separator at
GANIL and the technique of implantation into silicon detectors
for spectroscopic studies \cite{Borrel}. Later more detailed work was devoted
to $^{43}$Cr at this laboratory and all results are summarized in a
recent paper by Dossat et al. \cite{Dossat}. The half-life
was found to be $T_{1/2} = (21.1 \pm 0.4)$~ms. Several $\beta$-delayed proton
branches as well as one $\beta$-delayed two-proton channel were
identified. The total probability for emission of protons in the decay
of $^{43}$Cr was determined to be 92.5(28)\% \cite{Dossat}. Such large
value is expected, since the $\beta$-daughter nucleus, $^{43}$V is weakly
bound --- according to the systematics of Audi et al. its proton
separation energy is predicted to be $S_p = 190(230)$~keV \cite{AudiWapstra}.
From the same systematics, the decay energy of $^{43}$Cr is estimated
to be $Q_{EC} = 15.9 \pm 0.2$~MeV. In the work of Dossat et al., the total
branching ratio for emission of protons was deduced from the decay-time analysis
of the total charged-particle spectrum, with somewhat arbitrary assumptions
on the threshold energy and a contribution from $\beta$ particles \cite{Dossat}.
Such a procedure was necessary because not all proton-emission channels could be
identified as lines in the spectrum. For the same reason, no absolute
branching ratios could be determined separately for $\beta$-\emph{p} and
$\beta$-\emph{2p} channels. No evidence for the $\beta$-\emph{3p} transitions
was observed.

When $^{45}$Fe undergoes the \emph{2p} radioactivity it transforms into $^{43}$Cr,
thus the decays of the latter can be observed after implantation of the former.
Indeed, in our \emph{2p} decay study of $^{45}$Fe we could identify $\beta$-delayed
proton decay of $^{43}$Cr following the \emph{2p} emission from $^{45}$Fe \cite{Miernik3}.
We recorded 17 events of $\beta$-\emph{p} and
two events of $\beta$-\emph{2p} channels. These numbers were
found to be consistent with time observation windows imposed on
$^{45}$Fe implantation events, with the half-life of the $^{43}$Cr, and with
the known branching ratios \cite{Dossat}. However, in the reaction used
to produce the ions of $^{45}$Fe, also the $^{43}$Cr ions are produced. Since the
mass-to-charge ratio for these two nuclei is similar, and the production
cross section for the $^{43}$Cr is much larger, the latter ions are
transmitted to a detection apparatus in large quantity offering the possibility
to study their decays on their own. In this paper, we focus on decay
studies of such primarily produced $^{43}$Cr ions implanted in our detector.
We note, that in a different study of \emph{2p} radioactivity of $^{45}$Fe performed
by Giovinazzo et al. a large number of $^{43}$Cr was implanted as well
into a TPC detector used for decay spectroscopy \cite{Giovinazzo}.
Their detection of $\beta$-\emph{2p} events represents the first direct
observation of the two protons emitted after $\beta$ decay \cite{Giovinazzo}.
However, no detailed branching ratios were extracted, and no evidence for
the $\beta$-\emph{3p} decay was found in this work.

Our experiment was performed at the National Superconducting Cyclotron
Laboratory at Michigan State University, East Lansing, USA. The experimental
details were given already in Refs.\cite{Miernik1,Miernik3}, here we recall
only the most important points. The ions of interest were produced in the
fragmentation reaction of a $^{58}$Ni beam at 161 MeV/nucleon, with average
intensity of 15 pnA, impinging on a 800 mg/cm$^2$ thick natural
nickel target. The products were selected using the A1900 fragment
separator \cite{A1900} and identified in-flight by using time-of-flight
(TOF) and energy-loss ($\Delta E$) information for each ion.
The TOF was measured between a plastic scintillator located in the middle
focal plane of the A1900 separator and a thin silicon detector mounted at
the end of the beam-line. The silicon detector also provided the $\Delta E$ signal.
Identified ions were slowed down in an aluminium foil and stopped inside the active
volume of the OTPC detector. The acquisition system was triggered
only by those ions for which the TOF value exceeded a certain limit.
This limit was selected in a way to accept all ions of $^{45}$Fe and
a small part of the $^{43}$Cr ions. The rate of triggers due to $^{43}$Cr
ions was about 8 events per minute. The identification plot of all ions
coming to to the counting station was presented in Ref.\cite{Miernik3}. Here,
in Figure 1 we show the identification spectrum of those ions which triggered
the OTPC acquisition system. The total number of $^{43}$Cr events, selected for
the further analysis by the gate shown in Figure 1, was about 40000.

\begin{figure}[ht]
\begin{center}
    \includegraphics[width=1.0\linewidth]{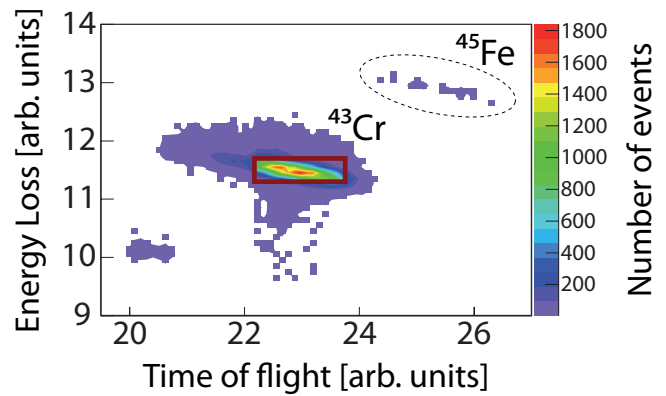}
    \caption{(Color online) Identification spectrum of ions accepted by
             the OTPC acquisition system. The gate used to select $^{43}$Cr ions
             for the further analysis is shown by a brown rectangle. }
\end{center}
\end{figure}

The principles of operation of the OTPC detector are
following. Ions and their charged decay products were stopped in a volume
of $20\times 20\times 42$~cm$^3$ filled with a counting gas at
atmospheric pressure. A mixture of helium (66\%),
argon (32\%), nitrogen (1\%), and methane (1\%) was used.
The primary ionization electrons drift
in a uniform electric field, with a velocity of 0.97~cm/$\mu$s, towards
a double-stage amplification structure formed by parallel-mesh
flat electrodes. In the second multiplication stage, emission of
UV photons occurs. After conversion of their wavelength to the
visual range by a thin luminescent foil, these photons were
recorded by a CCD camera and by a photomultiplier tube (PMT).
The camera image represents the projection of  particles' tracks
on the luminescent foil.
The signals from the PMT were sampled with a 50 MHz frequency
by a digital oscilloscope
yielding the time dependence of the total light intensity.
This provides timing information of events and additionally on
the drift-time which is related to the position along the axis
normal to the image plane. By changing the potential of an
auxiliary gating electrode, the chamber can be switched between
a low sensitivity mode in which tracks of highly ionizing heavy
ions can be recorded, and a high sensitivity mode used to detect
light particles emitted during the decay.

For each event, the corresponding CCD image and the
PMT time profile, assigned unambiguously to the accepted identified
ion, were stored on a computer hard disk.
The trigger signal was also used to switch the OTPC to the high
sensitivity mode and to turn the beam off for a period of 50~ms to
prevent other ions from entering the detector while waiting for
the decay of the stopped ion. While the sampling of the PMT signal was
always started by the trigger, two different modes of camera operation
were used. In the first, the \emph{asynchronous} mode, the CCD was
running continuously so that the frames were not correlated with the triggers.
The camera exposure time in this mode was 30~ms per frame.
Upon arrival of the trigger, the
current frame was validated and read-out after the exposure, while
neighboring frames were discarded. In this mode the track of the incoming
ion was recorded. Since the arrival of the ion occurs randomly
within the frame exposure time, the time left for the detection of the
decay is different in each case. To avoid this disadvantage, in the second
\emph{synchronous} mode, the CCD exposure was started by the
trigger signal. The exposure time in this mode was 24~ms.
Since the camera could start recording an image only
after about 100 $\mu$s after the trigger, the track of the incoming
heavy ion could not be seen in this mode. On the other hand, the
time window for the decay detection was the same for each ion and
equal to the camera exposure time. Out of about 40000 selected $^{43}$Cr
triggers, about 30000 were collected in the synchronous mode, and
about 10000 in the asynchronous one.
\begin{figure*}[th]
\begin{center}
    \includegraphics[width=0.95\linewidth]{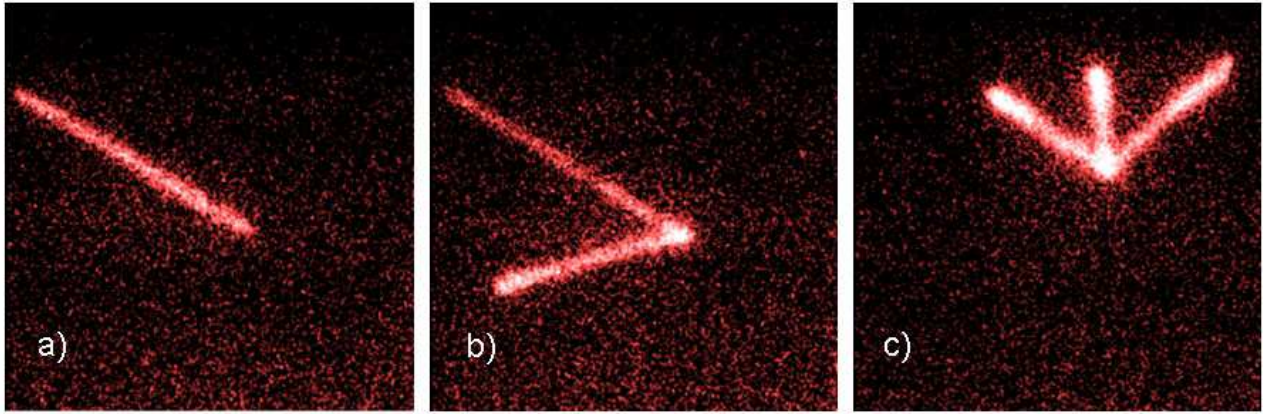}
    \caption{Example CCD images of $^{43}$Cr decays by $\beta$-delayed
    proton(s) emission, recorded in the synchronous mode in which the track of
    an incoming ion is not seen. The cases of emission of a single
    proton (a), two protons (b) and three protons (c) are shown.
    The ionizing power of the positron is too small to see its trace.}
\end{center}
\end{figure*}
Since the range distribution of products selected by A1900 separator
was larger than the thickness of the OTPC detector,
not all ions which triggered the acquisition system were actually stopped
within the active part of the chamber. Moreover, the experimental
conditions were optimized for ions of $^{45}$Fe which had slightly
different range profile. In addition, because the exposure time, also chosen
for the decay of $^{45}$Fe, was comparable with the half-life of $^{43}$Cr,
a significant fraction of the stopped $^{43}$Cr ions decayed after the exposure
was closed. Finally, even if an ion is properly stopped and does decay
within the observation time, there is a probability that the decay is
not accompanied by $\beta$-delayed protons. In such case only the track
of the incoming ion will be seen in the asynchronous mode, and nothing will
be seen in the synchronous mode, because the ionization power of a
positron is too small to leave a trace in the detector.

Among decays collected in the synchronous mode, we have observed
11502 events with emission of a single proton, 1010 events with simultaneous
emission of two protons and in 12 events three protons were seen to
emerge simultaneously from the decay vertex. Example events from
these three categories are shown in Figure 2. The clear and unambiguous
images of three-proton emission, like the shown in Figure 2c, establish
the $^{43}$Cr as the second nucleus, after $^{45}$Fe \cite{Miernik1},
in which such decay channel is identified. The numbers of events
given above yield the relative branching ratios of 91.8(3)\%, 8.1(3)\%,
and 0.096(30)\% for the $\beta$-delayed one-, two-, and three-proton
emission, respectively. Decay times of events in the synchronous mode,
recorded by the PMT tube, allow to extract the half-life of the $^{43}$Cr.
The histogram of decay times for all events of $\beta$-\emph{1p}
emission observed in this mode is shown in Figure 3.
The least-square fit to the data yielded the
half-life value of $T_{1/2} = 20.6 \pm 0.9$~ms, in perfect agreement
with the best literature value \cite{Dossat}.

\begin{figure}[th]
\begin{center}
    \includegraphics[width=1.0\linewidth]{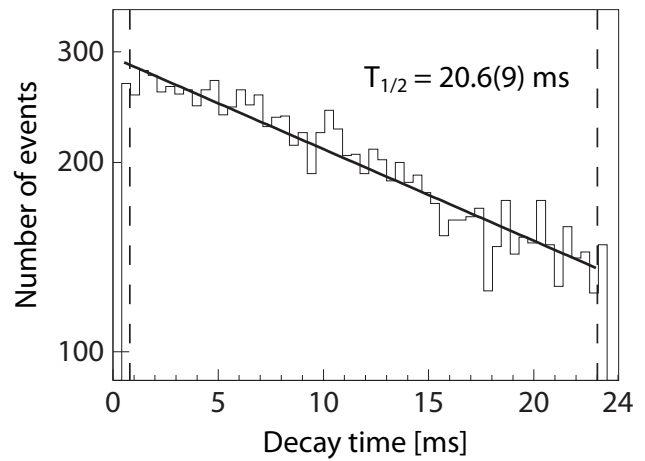}
    \caption{The histogram shows the decay time distribution
              of $^{43}$Cr events collected in the synchronous mode.
              The solid line shows the result of the least-square fit to the
              data points located between the dashed lines.}
\end{center}
\end{figure}

In the data collected in the asynchronous mode, we found 1663 events of
single proton emission, 117 events of two proton emission, and one event
of three proton emission, which is consistent with relative branchings
deduced from the synchronous mode. In all these events the track of the
incoming $^{43}$Cr ion is seen. The $\beta$-\emph{3p} event observed in
this mode is shown in Figure 4. In addition, in 4621 events we could
see only the heavy ion track, with no decay signal. These events represent
cases where either the decay occurred after the exposure time, or the
decay without proton emission took place. Since we know the half-life
of $^{43}$Cr and the time between the implantation and the end of the exposure
for each event (the decay time window), we can calculate the probability
that the decay with no protons did occur during such event. Thus, the analysis
of events in the asynchronous mode allows determination of the absolute
branching ratios for all decay channels.
\begin{figure}[th]
\begin{center}
    \includegraphics[width=1.0\linewidth]{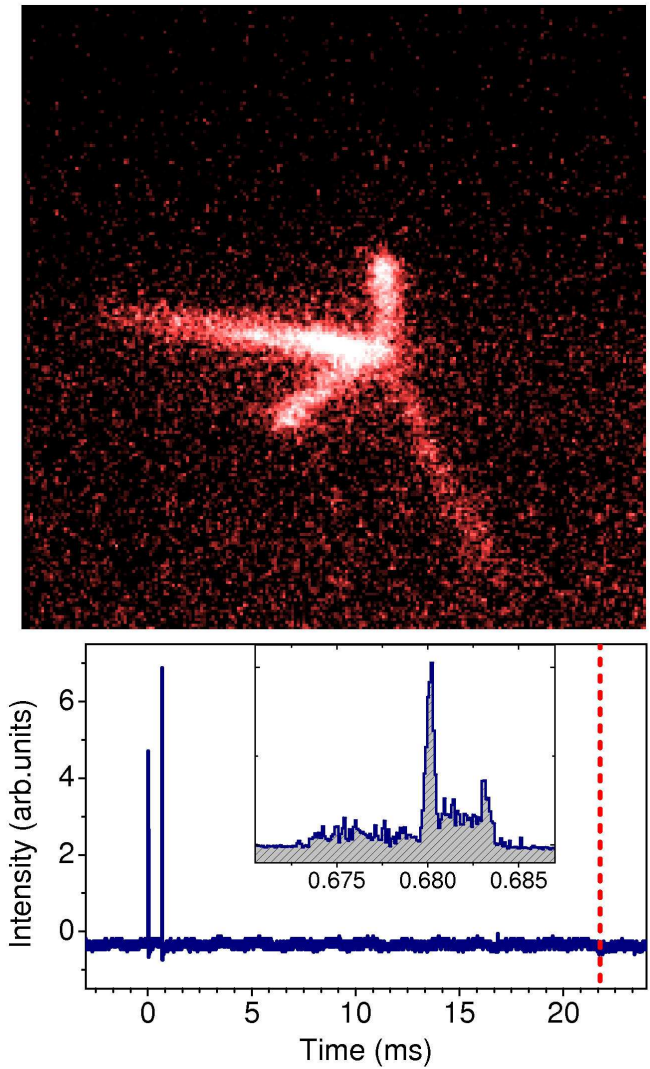}
    \caption{(Color online) An event of $\beta$-delayed three-proton emission from $^{43}$Cr
             recorded in the asynchronous mode. Top: the CCD image on which a track of a chromium ion
             entering the detector from the left can be seen. Bottom: the time profile of the
             total light intensity measured by the PMT. The dashed line indicates the end
             of the current CCD exposure frame. The clock is started by the implantation
             of the $^{43}$Cr ion. The decay event occurred 0.68~ms after the implantation
             and 21.1~ms before the end of the CCD frame. In the inset, the magnified part is
             plotted, showing in detail the decay event. }
\end{center}
\end{figure}

If the probability for the $\beta$ decay with no emission of delayed
protons is $b_0$, then the likelihood function, defined as the product
of probabilities for all events, can be written as

\begin{displaymath}
    \mathcal{L} = \prod\limits_{i=1}^{N_p} [(1 - b_0)\,(1 - e^{\lambda \tau_i})]\,
                      \prod\limits_{j=1}^{N_0} [e^{-\lambda \tau_j} + b_0 (1 - e^{-\lambda \tau_j} )],
\end{displaymath}
where the first product goes over all $N_p$ decay events with at least one proton emitted,
the second products includes all $N_0$ events where no delayed protons were seen,
$\lambda$ is the decay constant of $^{43}$Cr, and $\tau_i$ is the decay time
window in the $i$-th event. According to the maximum likelihood method, the value of
$b_0$ can be determined by finding the maximum value of the function $ \mathcal{L}$.
This procedure applied to all events accumulated in the asynchronous mode in which the
ion was clearly stopped within the active volume yielded the value $b_0 = 0.12 \pm 0.04$.
Combining this value with the relative probabilities discussed earlier, we
arrive at the absolute branching ratios for the various decay branches of $^{43}$Cr
which are shown in Table 1. For comparison, the results from Ref.\cite{Dossat}
are also given in Table 1.
The values for the one- and two-proton emission were taken from Table 9 in Ref.\cite{Dossat},
they represent only those transitions which could be identified as peaks in the
proton spectrum. Many decay channels are missed in that way which is reflected
by the difference between the determined total probability for emission of
protons (92.5\%) and the sum of identified one- and two-proton contributions.
However, the branching ratio for $\beta$-\emph{2p} channel, attributed in Ref.\cite{Dossat}
to a transition from the IAS state in $^{43}$V to the ground state of $^{41}$Sc,
agrees within error bars with the value determined by us. This may indicate that
this transition is the only one with the $\beta$-delayed two-proton emission, or
at least it dominates this decay channel. Also the two branching ratios for
decays with no protons, given in Table 1, agree within experimental accuracy.
Thus, the missed channels, i.e. not identified as lines in the proton spectrum
in Ref.\cite{Dossat}, are predominantly of the $\beta$-\emph{1p} type.

The $\beta$-delayed three proton channel is found to be very weak, with the branching
below the $10^{-3}$ level. This may explain why this decay mode was not
identified in previous experiments on $^{43}$Cr. Its observation in the present
work illustrates the extreme sensitivity of the OTPC detector. In principle
a single image would suffice as an evidence for such a rare decay mode. We note
also that in this exotic transition a proton-drip line nucleus ($^{43}$Cr)
transforms into a stable and doubly-magic one ($^{40}$Ca).

\begin{table}
\caption{The absolute branching ratios for the decay channels of $^{43}$Cr
         with various number of $\beta$-delayed protons.
         The values with a star ($^*$) were determined
         in Ref.\cite{Dossat} for channels which could be identified
         as lines in the proton spectrum. }
\vspace{0.5\baselineskip}
\begin{centering}
\begin{tabular}{ccc}
\hline
   \multicolumn{1}{c} {Number} &
   \multicolumn{2}{c} {Branching ratio} \\
   \multicolumn{1}{c} {of protons } &
   \multicolumn{1}{c} {This work} &
   \multicolumn{1}{c} {Ref.\cite{Dossat}} \\
\hline
   0 &  0.12(4)   & 0.075(3)   \\
   1 &  0.81(4)   & 0.28(1)$^*$    \\
   2 &  0.071(4)  & 0.056(7)$^*$   \\
   3 &  0.0008(3) &    -       \\
\hline
\end{tabular}
\end{centering}
\end{table}

The relatively large branching ratio for transitions without $\beta$-delayed protons
is surprising. It could be expected that it results from strong feeding of the ground state
of $^{43}$V and of its first excited states lying sufficiently close to the proton
separation energy for proton emission being suppressed.
However, the assumption of isospin symmetry
and the known $\beta$ feedings in the mirror decay of $^{43}$K to $^{43}$Ca, indicate
that such an explanation is very unlikely. The ground-state to ground-state
transition in $^{43}$K is a first forbidden unique ($3/2^+ \rightarrow 7/2^-$) with
$\log ft =9.71$ \cite{Warburton}. Such a $\log ft$ value would correspond to the
decay of $^{43}$Cr to the ground state of $^{43}$V with the probability of 0.12\% only.
The strongest allowed transition in the decay of $^{43}$K to the 990 keV
state in $^{43}$Ca has the $\log ft =5.57$ which would yield a branching of 2\%
in the decay of $^{43}$Cr. The summed feeding to the ground state and the first
four excited states in $^{43}$Ca, up to the 1394 keV energy, would yield less
than 3\% branching in the decay of $^{43}$Cr if the mirror symmetry were perfect.
Moreover, the $\beta^+$ transitions to the unbound states are observed to be
systematically weaker than the corresponding mirror $\beta^-$ channels \cite{Borge,Piechaczek},
which would rather reduce the expected feeding of states in $^{43}$V.
It seems plausible that in addition to possible isospin asymmetry effects and/or
a serious discrepancy in the proton separation energy in $^{43}$V, some proton-unbound
states fed in the decay of $^{43}$Cr have $\gamma$-widths much larger than the proton widths.
Such a case of $\gamma$ transitions winning a competition with the proton emission
was observed in the decay of $^{37}$Ca \cite{Trinder}. The same effect was
suggested as an explanation of an apparent reduction in the feeding of an
unbound state in the decay of $^{20}$Mg \cite{Piechaczek}.
However, to check whether the fast $\gamma$ transitions do explain the large branching
for decays with no $\beta$-delayed protons, a study invoking both high-resolution
$\gamma$ and proton spectroscopy would be required.

In summary, we have applied a new type of ionization chamber, in which the
optical imaging technique is used to record tracks of charged
particles, to the study of $^{43}$Cr $\beta$ decay.
Although the experimental conditions were optimized for the measurement of \emph{2p}
radioactivity of $^{45}$Fe \cite{Miernik3} performed during the same experiment,
a large number of $^{43}$Cr ions was implanted in the detector and events
associated with their $\beta ^+$-delayed proton emission could be clearly
identified. Among them, thirteen events of $\beta$-delayed three-proton
emission were recorded. Thus, $^{43}$Cr is established as the second case,
after $^{45}$Fe \cite{Miernik1}, where this exotic decay mode is observed.
In most cases the emitted protons had energies large
enough to escape from the detector's volume, making the detailed kinematical
characterization of the decay impossible. However, the relative, as well
as the absolute branching ratios for decays with different number of
delayed protons could be directly determined. A large branching ratio of ($12 \pm 4$)\%
was found for decays with no $\beta$-delayed protons, although
the feeding of proton-bound states in the one-proton daughter nucleus $^{43}$V
is expected to be very small by isospin symmetry arguments. Such large value
could result from the proton separation energy in $^{43}$V being larger than
expected from systematics \cite{AudiWapstra}, from substantial isospin asymmetry
effects, and from large $\gamma$-widths of some proton-unbound states in $^{43}$V.

This work was supported by a grant from the Polish Ministry of Science
and Higher Education number 1 P03B 138 30, the U.S. National Science Foundation
under grant number PHY-06-06007,
and the U.S. Department of Energy under contracts DE-FG02-96ER40983,
DEFC03-03NA00143, and DOE-AC05-00OR22725. A.K. acknowledges the support
from the Foundation for Polish Science.

\end{document}